# Quantum Resistance in Multilayer Graphene-BiFeO$_3$ Memristor for Brain-Inspired Computing

Suman Roy[1,2], Priyanka Sahu[1,3], Subhabrata Das[6], Sameer Kumar Mallik[4], Susmita Jana[7], Alok Kumar[1,2], Himadri Nandan Mohanty[5], Kaushik Ghosh[6], B.R.K. Nanda[7] and Satyaprakash Sahoo[1,2,*]

[1]Laboratory for Low Dimensional Materials, Institute of Physics, Bhubaneswar-751005, India
[2]Homi Bhabha National Institute, Training School Complex, Anushakti Nagar, Mumbai-400094, India
[3]School of Physics, Sambalpur University, Jyoti Vihar, Burla, Odisha 768019, India
[4]Department of Microtechnology and Nanoscience, Chalmers University of Technology, Göteborg SE-41296, Sweden
[5]Centre de Nanosciences et de Nanotechnologies, CNRS, Université Paris-Saclay, 91120, Palaiseau, France
[6]Quantum Materials & Devices Unit, Institute of Nano Science and Technology, Knowledge City-Sector 81, Mohali 140306, India
[7]Condensed Matter Theory and Computational Lab, Department of Physics, Indian Institute of Technology Madras, Chennai 600036, India

**Abstract**
In the era of big data and the Internet of Things, quantum-level control of conductance states offers a promising route toward high-density data storage and brain-inspired neuromorphic computing. Although quantum conductance (QC) phenomena have been demonstrated in various metal oxide memristors, achieving reliable and precise control over quantized states remains in its infancy. Here, we demonstrate bidirectional quantum conductance states in multifunctional BiFeO$_3$ (BFO) perovskite memristors integrated with multilayer-graphene contacts, enabling higher-order tunability and revealing the potential of perovskite-2D heterostructures for quantum-engineered memory and computing devices. XPS analysis provides detailed insights into oxygen vacancy dynamics in BFO, whereas first-principles density functional theory calculations clearly reveal a strong localized electric field at the graphene-BFO interface. Our devices exhibit current-controlled higher-order QC transitions facilitated by quantum point contact formation, giving rise to quantized conductance states during both SET and RESET processes. Time-lag correlation maps quantify the stochastic evolution of QC states under dynamic voltage-pulse tuning schemes. Notably, the quantized conductance states effectively emulate synaptic potentiation and depression, enabling precise weight modulation for high-accuracy image and digit recognition in convolutional neural networks. These findings establish perovskite-2D heterostructures as promising candidates for QC-driven resistive switching and demonstrate their potential for developing controllable quantum memristors.

*Corresponding author: sahoo@iopb.res.in

# Introduction

The continuous evolution of electronic devices toward higher performance, lower power consumption and enhanced functionality has placed memristive technologies at the forefront of research in neuromorphic computing and non-volatile memory (NVM) applications.[1,2] Metal-oxide resistive random-access memory (RRAM) stands out among next-generation NVM technologies due to its high storage density, notable scalability, high endurance, streamlined device architecture, rapid switching capabilities and seamless integration with traditional complementary metal-oxide-semiconductor (CMOS) technology.[3–5] Memristive devices are commonly structured as two-terminal systems in a metal-insulator-metal (MIM) configuration, where an active layer is placed between two metal electrodes.[6,7] Their operation is governed by the resistance state of the device, which evolves according to the history of the applied voltage and current. Among the various working mechanisms used to develop these devices, redox-based memristors-nanoscale electrochemical systems that integrate ionic and electronic processes, have gained significant attention.[8,9] Notably, in these cases, the device behaviour is largely governed by atomic reconfiguration driven by nano-ionic species.[8,10] In electrochemical metallization memory (ECM) devices, resistance modulation occurs through metal cation transport within the active material by forming an atomic filament that bridges the electrodes, resulting in a transition from a high-resistance state (HRS) to a low-resistance state (LRS) of the device. In contrast, valence change memory (VCM) devices rely on oxygen anion movement across the metal-oxide active layer, forming an oxygen vacancy-rich conductive filament (CF). Here, device functionality is regulated by ion dynamics, while electrons assess the resistance state internally.[11]

Interestingly, when the thickness of the CF in memristive devices approaches the atomic scale, quantum conductance (QC) is often observed in the electrical transport characteristics. In the QC phenomenon, electron transport through the CF adheres to the Landauer criterion, resulting in ballistic conduction. This results in the formation of a quantum point contact (QPC), where conductivity is quantized in units of the fundamental conductance, $G_0$ $(=\frac{2e^2}{h})$, $e$ is the electron charge and $h$ is the Planck constant) (77.5 µs or 12.91 kΩ$^{-1}$) at room temperature.[12,13] [14] The engineering of the QC states in memristor devices could offer an additional degree of freedom for exploring neuromorphic computing through tuneable quantized conductance states.

Significant progress has been made in understanding resistive switching (RS) and memristive computing architectures, harnessing quantum effects like ballistic transport in these devices remains a developing frontier, offering new avenues for practical applications. Despite notable advancements, achieving stable and controllable QC in memristive devices remains challenging, limiting the broader application of these devices in scalable technologies.[15,16] Studies reveal that maintaining higher QC states is often hindered by instability and unpredictability, largely due to factors such as random telegraph noise caused by atomic thermal motion within the CF and nonlinear QC effects arising from CF broadening.[17–19] These random fluctuations in conductance degrade the device's performance and restrict its scalability. Additionally, QC is sensitive to various external and internal influences, including Joule heating, electric field gradients, the Gibbs-Thomson effect, mechanical stress from surrounding insulating layers, and environmental factors.[20,21] These factors introduce further fluctuations, making it difficult to achieve desirable stable QC states.

To address these issues, researchers have been increasingly focused on regulating and stabilising QC in memristors and their potential use beyond Moore's law.[15,18,22,23] This interest has led to the development of electrical control techniques that aim to refine and stabilise the QC states. For example, controlling the compliance current (CC) helps to limit the maximum current flowing through the device, thus reducing unwanted thermal effects.[24] Adjusting the stopping voltage and varying the voltage sweep rate are also methods used to influence the formation and stability of the atomically thin CF.[22,25,26] Similarly, using different current sweeping modes allows fine control over the QC state transitions, aiming to achieve reliable quantized conductance even under challenging conditions.[16] In addition to these electrical tuning parameters, the choice of metal contact plays a crucial role in stabilising and modulating QC states, as the interfacial energy landscape and carrier injection properties strongly affect filament dynamics at the atomic scale.[11] Collectively, these techniques show promise in overcoming the inherent limitations of QC in memristors, potentially paving the way for their integration into future electronic applications.

Considering the significant influence of the switching layer on QC, a variety of materials have been explored to attain higher-order QC states. These include metal oxides like $HfO_x$,[27] $TiO_x$,[24] and $ZnO$[18]; ionic conductors such as $GeS_2$,[28] $Ag_2S$,[14] and $AgI$[29]; as well as organic compounds, including poly(3-hexylthiophene): [6,6]-phenyl-C61-butyric acid methyl ester (P3HT: PCBM),[30] two-dimensional (2D) materials like $MoS_2$,[31] h-BN[32] and polyethylene oxide systems.[33] Alongside the active switching medium, different metal contacts such as Pt, Ag, Au, Nb, and Cu have also been employed to realise QC behaviour, highlighting the critical role of electrode interfaces.[15,18,24,27,32] However, memristors based on 2D materials combined with various complex oxides have garnered significant attention due to their unique electrical properties and potential for multifunctionality. Recently, Xie et al. have demonstrated the tunability of QC in h-BN by carefully engineering graphene edge contact, where the atomically thin edge contact confines the active switching area to only a few nanometers.[34] This localization enables the isolation of a single conductive filament, resulting in a clean single-step quantized conductance transition. While this work highlights the ability of graphene to act as an atomically sharp injector for quantum transport, the integration of graphene with other technologically active materials, such as novel metal oxides and complex organic materials systems for quantum conductance control, remains unexplored. In our previous study, we have demonstrated pulse tuneable higher-order QC states in oxygen vacancy-engineered Ag/TiO$_{2-\Delta x}$/Pt memristors; however, the QC behaviour was observed only during the RESET process, limiting the bidirectional control of quantum states.[24]

In this work, we report the demonstration of bidirectional quantum conductance in multilayer graphene (ML-Gr) contacted multifunctional Bismuth Ferrite (BiFeO$_3$) perovskite-based memristor. The ML-Gr/BiFeO$_3$ (BFO)/FTO vertical memristor exhibits CC controlled and pulse-tuneable QC states extending to high-order quantum levels during both SET and RESET processes. The defect-rich ferroelectric oxide supports the formation of multiple conductive filaments or variable-width atomic channels. In addition to that, first-principles calculations validate that graphene defects act as charge-trapping centres, generating interfacial electric field enhancement. The experimentally obtained QC states serve as synaptic weights in convolutional neural network (CNN) simulations for colourful image and digit pattern recognition performance. This work demonstrates that graphene contact engineering, which enables atomic-scale confinement in crystalline 2D systems, can be generalized to metal oxide memristors to stabilize and control multilevel quantum states.

# Results and Discussion:

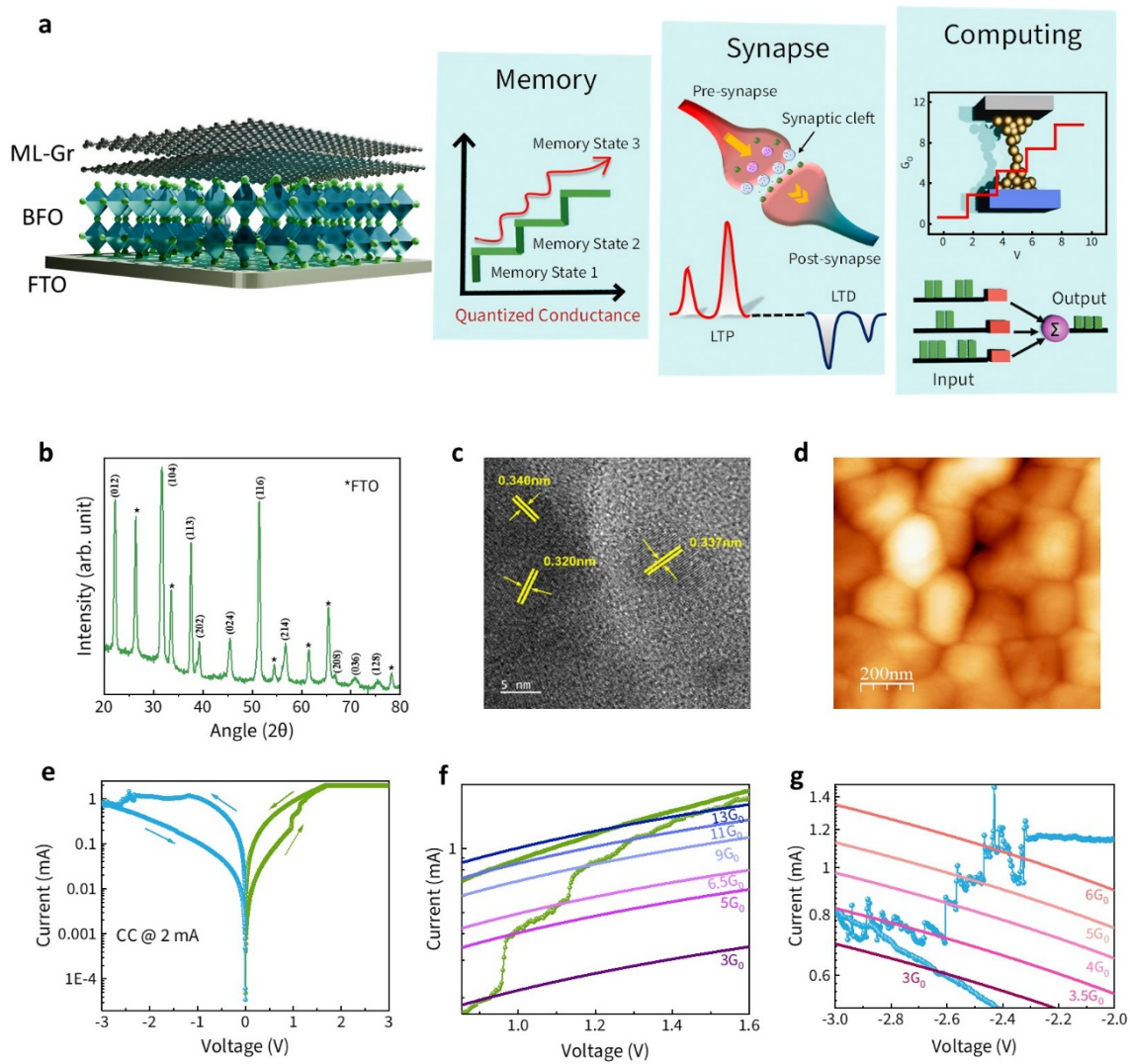

**Figure 1.** *Characterization and switching behaviour. (a) Left: Schematic of the Multi-layer Graphene (ML-Gr)/BFO/FTO memristor device. Right: Its use in quantum point-contact memory, and biologically inspired synapses, and pattern-recognition tasks using unconventional computing architectures. (b) XRD pattern of the BFO/FTO thin film. (c) HRTEM and (d) AFM image of the BFO thin film. (e) A typical DC resistive switching cycle of the memristor. (f) and (g) present the zoomed-in views of the QC states during the SET and RESET processes, respectively, showing discrete steps fitted to integer and half-integer multiples of $G_0$.*

**The ML-Gr/BFO/FTO Memristor Architecture and Material Characterization.** Figure 1a presents the schematic of the ML-Gr/BFO/FTO memristor device architecture, highlighting its potential applications in quantum point-contact (QPC) memory, biologically inspired synapses, and pattern-recognition tasks using unconventional computing architectures. The XRD pattern of the synthesised BFO thin film is shown in Fig. 1b. It is clearly seen that it is well matched with JCPDS data 74-6717, having a polycrystalline nature of perovskite-type structure (α-BifeO$_3$) with rhombohedral space group R3C.[35] The structural characterization is further supplemented with Raman spectroscopy. Raman spectrum of BFO is depicted in Fig. S1a of the supplementary information (SI), which shows the vibrational modes of BFO & Graphene in the range from 60 to 3000 cm$^{-1}$. For pristine BFO film, 4A1 and 5E modes are observed at 131, 168, 212, 488, 272, 345, 416, 532 and 590 cm$^{-1}$, which is in good agreement with the literature.[35] The A1 modes correspond to longitudinal Bi-O bonds, while transverse Fe-O vibration bonds contribute to the E modes. Figure S1b of the supplementary information shows the optical absorbance spectrum of the as-synthesised BFO thin film. The spectrum typically exhibits a strong absorption edge in the visible region. We further analyse the absorption spectra to estimate the band gap of BFO. The inset shows the optical tauc plot of the UV-Vis spectra, and the band gap is found to be 2.32 eV. HRTEM micrographs of the BFO sample extracted (a small portion of the sol-gel synthesised thin film is scratched from the substrate and transferred onto the TEM grid to carry out HRTEM studies) from the thin film is shown in Fig. 1c. Further analysis with the HRTEM of individual particles confirms clear evidence of the ultra-fine single crystal structure, showing an interplanar spacing of d = 3.2 Å, corresponding to (012) crystal planes. In addition to that, the selected area electron diffraction (SAED) shows a bright diffraction spot with diffuse rings superimposed with spots, which confirms the formation of a polycrystalline structure, provided in Fig. S1c of the SI. The X-ray energy dispersive spectroscopy (EDS) spectrum showcased in Fig. S2a of the SI displays prominent peaks corresponding to bismuth (Bi), iron (Fe), and oxygen (O), which fairly agrees with the expected stoichiometry of the BFO. Additionally, elemental mapping as depicted in Fig. S2b of the SI reveals a uniform distribution of Bi, Fe and O across the film surface, supporting good compositional homogeneity. The AFM studies are further carried out to understand surface morphology, grain size and roughness of the thin film. From the AFM images as shown in Fig. 1d, one can notice the presence of small grains of size about 100-200 nm, which are quite common in sol-gel grown BFO. [36] AFM topography reveals a dense, uniform surface morphology with an average RMS roughness of ~ 9 nm over a 1 × 1 µm² area, indicating well-connected grains suitable for reliable electrical performance.

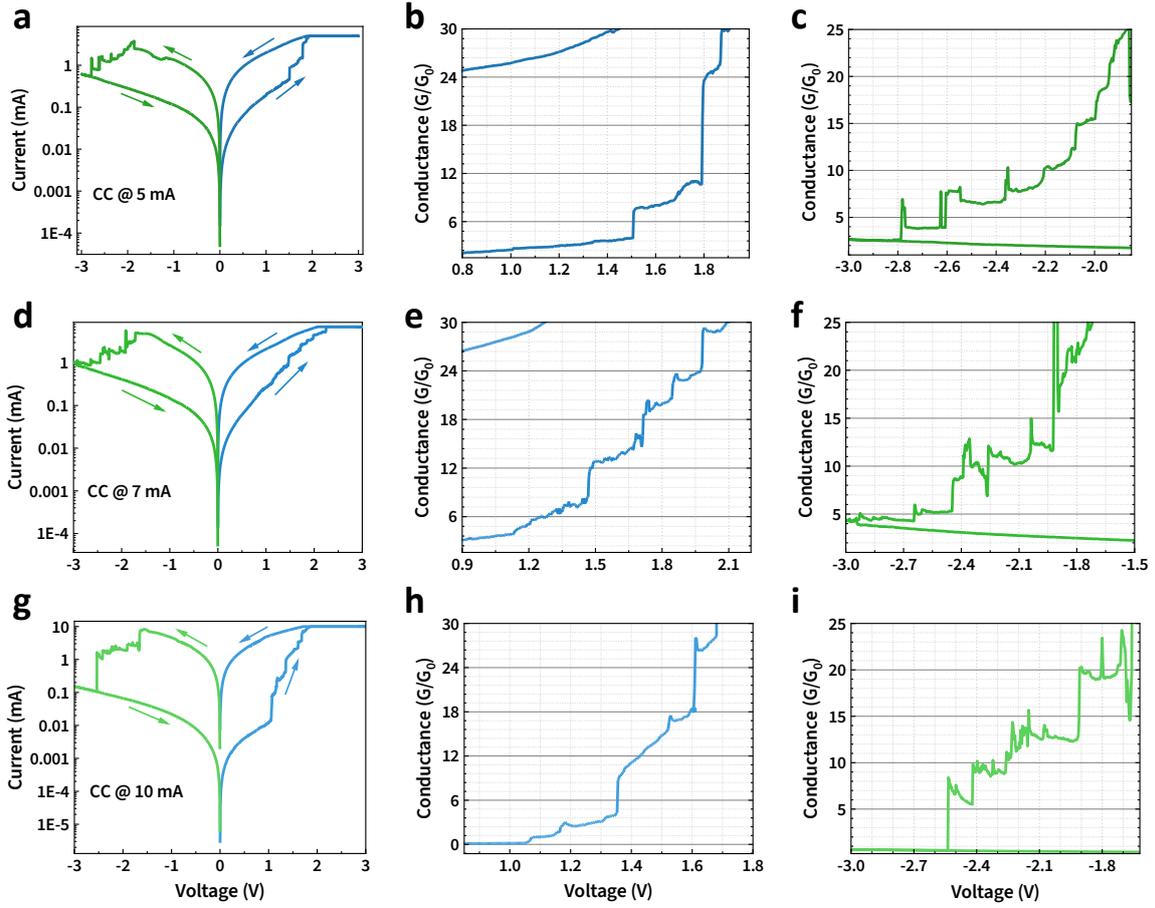

**Figure 2.** *Compliance Current dependent Quantized conductance states in sol-gel grown BFO-based memristors. A typical DC resistive switching cycle of the memristor at the compliance current 5, 7 and 10 mA is displayed in (a), (d) and (g). The corresponding conductance vs voltage characteristics during SET are represented in (b), (e), and (h), and during RESET in (c), (f), and (i), illustrating both integer and half-integer multiples of $G_0$.*

**Quantum Conductance Detection.** Quantum conductance switching, characterized by discrete conductance changes at quantized levels, represents a significant paradigm shift from the conventional continuous resistive switching behaviour, which is usually observed in memristors. To investigate the evolution of QC states of ML-Gr/BFO/FTO memristor, direct current (DC) I-V measurements at ambient conditions are carried out by sweeping the voltage at the top ML-Gr electrode while the bottom FTO electrode is grounded. In this study, at a low compliance current (CC) of 2 $mA$, the device exhibits non-volatile switching behaviour, which signifies the formation of a conducting filament due to the presence of oxygen vacancies ($O_v$) in the BFO active layer. During the transition from high resistance state (HRS) to low resistance state (LRS), some of the I-V characteristics of the memristor undergo small but distinct discrete steps instead of a conventional high sharp jump in the current, as shown in Fig. 1f at around $1\ V$ and $-2\ V$, respectively. In the I-V characteristics (with small jumps), $V_{SET}$ and $V_{RESET}$ are defined as the first current jump during the SET and RESET process, respectively. It may be noted that, filamentary conduction path is nucleated at $V_{SET}$, and initiation of the rupturing of the filament occurs at $V_{RESET}$. In literature, similar small jumps in I-V characteristics have been reported for $HfO_x$, $TaO_x$, and ZnO based memristors.[18,19,27] The number of such jumps is found to depend on the type of materials and electrode used, and the dimension of the material. The numbers of such jumps are related to the different QC states of the device. To represent these QC states more clearly, the electrical conductance, G, has been determined from the ratio of current to voltage, $G = I/V$. These conductance values are expressed in units of the quantum conductance, $G_0$. Multiple abrupt conductance transitions can be clearly observed, each succeeded by a distinct, steady conductance plateau as depicted in Fig. 1f and Fig. 1g for the SET and RESET process, respectively. It is observed that the current steps are well fitted to both integer ($nG_0$, where $n = 1, 2, 3..., 10$), half-integer ($mG_0$, where $m = 0.5, 1.5, 2.5, 3.5, 4.5$) and fractional quantum conductance changes. However, these fractional QC states are random and not very controllable within the applied CC. This suggests the conductive filament could be one or a few atoms thick, behaving as a quantum wire with one or several conducting channels. Stability of these states is verified through 50 consecutive I-V cycles, as shown in Fig. S3 of the SI, demonstrating reproducibility of the result.

**Current-Limited Quantum Conductance Phenomena.** The quantum conductance, $nG_0$, indicates that there are many intermediate states where the CF is a nano-sized filament with properties similar to a quantum wire.[37] The formation of these nano-sized filaments depends upon the geometry of the device and DC electrical measurement parameters. To further understand the growth dynamics of filaments and RS behaviour, I–V measurements are carried out over a spectrum of CC. Prior reports indicate that higher CC values promote more robust filament formation between electrodes in solid polymer electrolyte devices.[38] These studies imply, choosing an appropriate CC range is critical for controlling filament development and achieving quantum point contact (QPC) formation. In order to enable the controlled QC phenomenon, we have applied a DC sweep voltage ($0\ V \rightarrow 3\ V \rightarrow 0\ V \rightarrow -3\ V \rightarrow 0\ V$) under the optimized step voltage of $1\ mVs^{-1}$ with different CC as presented in Fig. 2. When a CC of 5 mA is applied under the same voltage sweep condition, we observed that the QC states with half integer $G_0$ and integer multiples appears stepwise in both SET and RESET process as shown in Fig. 2a. During SET process a sharp conductance change from $3G_0$ to $7G_0$ is observed at $1.5\ V$ followed by a plateau at $8.5G_0$ as depicted in Fig. 2b. Similarly, the conductance changes from $11G_0$ to $24G_0$ followed by a conductance plateau from $1.8\ V$ to $1.85\ V$. Then another conductance jump occurred and the conductance reached to CC. Likewise, the RESET process exhibited comparable G-V characteristics featuring stepwise conductance transitions as a function of voltage, as demonstrated in Fig. 2c. In the RESET process, the first conductance jump occurs from $22G_0$ to $19G_0$ followed by a plateau at $18G_0$. Then the conductance reduces gradually from $18G_0$ to $\sim 2.5G_0$. When CC is increased to $7\ mA$ (depicted in Fig. 2d), the QC states are observed in the range $\sim 4G_0$ to $28G_0$ during the SET process and from $23G_0$ to $\sim 4G_0$ in the RESET process, as replicated in Fig. 2e and Fig. 2f, respectively. When CC is further increased to $10\ mA$ (shown in Fig. 2g), higher conductance states are observed as shown in Fig. 2h during the SET process. At the RESET process, we observed a similar kind of QC behaviour, ranging from $24G_0$ to $\sim 0.5G_0$, represented in Fig. 2i. It may be noted that as the CC is increased, our devices demonstrate an expanded range of QC states extending to higher conductance values. The variability of the I-V characteristics across each CC over up to 100 cycles is shown in Fig. S4 in the SI. These observations provide strong evidence for QPC formation within the narrowest constrictions of the conductive pathways, demonstrating characteristic quantum conductance behaviour in the memristive devices. Half-integer quantum conductance values can originate from impurities absorbed on or within atomic chains, which alter the constriction geometry and modify the electronic band structure effects previously reported in metal monatomic chains with foreign absorbates.[25,39] The temporal stability of these conductance states is also verified by performing the retention measurement over extended periods exceeding $6 \times 10^3$ seconds, demonstrating excellent state preservation as illustrated in Fig. S5 (supplementary information). Furthermore, notable conductance oscillations were occasionally observed throughout the RESET process during different CC measurements, as demonstrated in Fig. S6 of the SI. It suggests that these fluctuations result from the formation and breakdown of nanoscale contacts within the conductive filaments. Zhu et al. reported clear conductance oscillations during RESET in Nb/ZnO/Pt resistive memory devices, linking the effect to the repeated formation and rupture of atomic-scale filaments.[18] Such conductance oscillations can be attributed to the dynamic balance between field-assisted filament formation and thermally-driven filament rupture within specific voltage and current operating windows, consistent with observations in similar kinds of device architectures.[40–42]

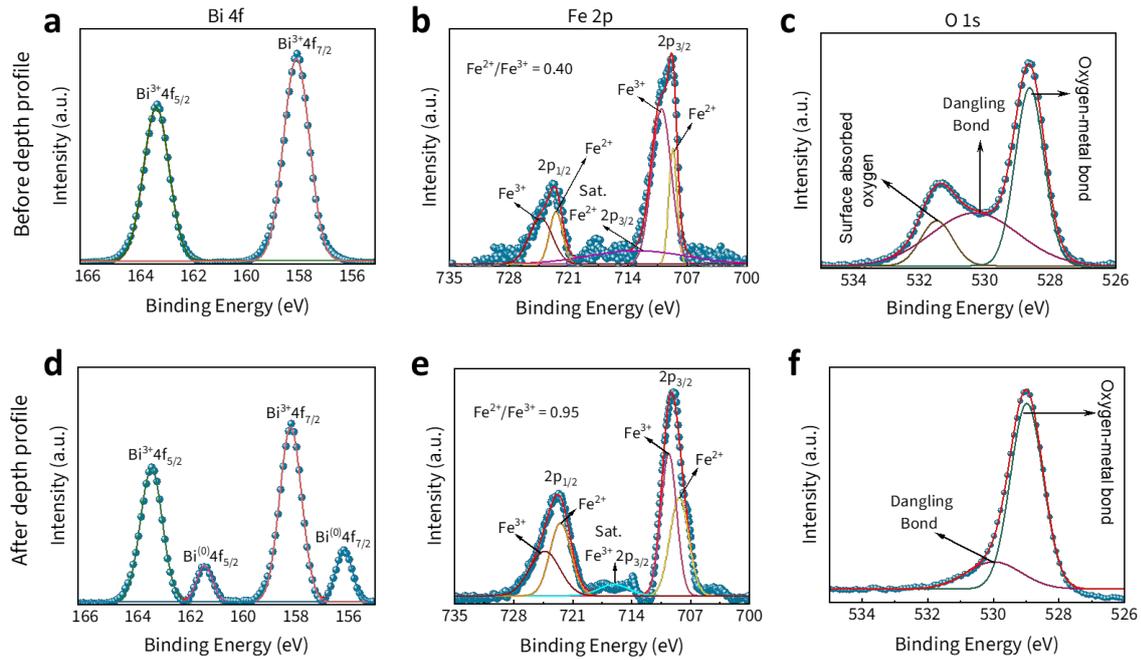

**Figure 3.** *XPS spectra of the BFO thin film before and after depth profiling. (a–c) show the high-resolution XPS spectra of Bi 4f, Fe 2p, and O 1s, revealing the presence of $Bi^{3+}$, mixed $Fe^{2+}/Fe^{3+}$ oxidation states with their corresponding ratio, as well as contributions from dangling (surface-adsorbed) and lattice oxygen bonds, respectively. After etching (d–f), the formation of metallic $Bi^{(0)}$ is observed along with an enhanced $Fe^{2+}/Fe^{3+}$ ratio. Additionally, a significant reduction in surface-adsorbed oxygen and the dominance of lattice oxygen peaks indicate the development of a vacancy-enriched subsurface region. These changes confirm the presence of oxygen vacancies, which are responsible for defect-mediated conduction and quantum transport in the ML-Gr/BFO/FTO memristor device.*

**Proposed Switching Mechanism Leading to Quantum Conductance States.** In our ML-Gr/BFO/FTO memristor configuration, resistive switching and the emergence of quantized conductance states are predominantly governed by the interplay between oxygen vacancy dynamics in the BFO layer and the electronic characteristics of the multi-layer graphene top electrode. By tailoring $O_v$ stoichiometry within the insulating layer can provide spatial controllability of CF geometry to effectively control the microscopic origin of QC states. Multi-layer graphene, employed as the top electrode, is known for its high electrical conductivity and tuneable work function. However, it also inherently contains atomic-scale edge defects such as vacancies or ripples introduced during the exfoliation process. These defects act as active sites that significantly influence the local electric field distribution and charge injection behaviour. Upon applying a positive bias to the graphene electrode, electrons are accumulated near the defect site, generating a strong localized electric field at the graphene-BFO interface. Our first principles calculation on the charge density localization around the graphene defect site with applied bias will be discussed later. Due to this electric field, oxygen vacancies tend to drift toward the FTO electrode. This migration promotes the formation of localized conductive filaments through the clustering of $O_v$ and the subsequent creation of reduced $Fe^{2+}$ and $Bi^{2+}$ states. However, the filament formation in this system is spatially non-uniform at the top interface (graphene/BFO), due to the defect-induced local field enhancement, a bulky or atomic filament structure tends to form.[43] This region experiences stronger localize ionic and electrical activity, resulting in a larger conduction cross-section and quasi-metallic conduction. Towards the bottom interface (BFO/FTO), the fluorine-doped tin oxide (FTO) electrode also acts as an $O_v$ reservoir, but it tends to stabilize vacancy distribution due to its relatively lower local electric field and higher oxygen affinity.[44] As a result, single or multiple narrow and well-confined conduction channels are formed at the Graphene/BFO interface. The confined top portion of the filament behaves as a quantum point contact, supporting ballistic electron transport through discrete conduction modes.[45] The conductance across this device quantizes in multiples of $G_0$, depending on the filament width and number of conducting channels.

Upon sweeping the applied voltage, the conductive paths continue to grow until one vacancy connects to the top graphene electrode, and the device switches from the HRS to the LRS. As the sweeping voltage increases, the multiple quantum point contacts are formed between top and bottom electrode and/or by branching the existing conducting filaments showcasing multiple QC states as shown in Fig. 1f. Similarly, when the polarity of the sweeping voltage is reversed, the number of quantum point contacts are reduced due to the gradually rupture of the conductive filaments and/or the reduction of conductive branches because of the drift of $O_v$ towards FTO electrode as shown in Fig. 1g. During SET and RESET processes, the modulation of filament cross-section and the number of point contacts, particularly near the graphene/BFO interface, results in multiple stable QC states which can be tuned by the compliance current levels. The increase in compliance current leads to the expansion of the bulk filament diameter and a higher number of quantum point contacts, accelerating oxygen ion diffusion and structural reconfiguration. This phenomenon is consistent with our observations, because of this, conductance states between $25G_0$-$40G_0$ are rarely observed in our study, indicating the device conductance states may not be "quantized" anymore, as the charge carrier scattering, transitioning the transport mechanism from ballistic to diffusive regime.[46] Therefore, we restrict our analysis

Up to $25G_0$. A schematic representation of the evolution of filament formation with increased CC is illustrated in Fig. S7 of SI.

To further elucidate the role of oxygen vacancies and their contribution to the resistive switching as well as quantum conductance behaviour, X-ray photoelectron spectroscopy (XPS) is performed on the surface and the ~10 nm depth-profiled regions of the BFO thin films. Figure 3 shows the high-resolution Bi 4f, Fe 2p, and O 1s spectra before (Figure 3a–c) and after (Figure 3d–f) Ar$^+$ etching. At the surface, the O 1s spectrum is deconvoluted into three components centred at 531.5 eV (surface-adsorbed oxygen), 530.3 eV (dangling bonds), and 528.6 eV (metal–oxygen bonds). Following depth profiling, the disappearance of the surface-adsorbed oxygen peak indicates the removal of superficial contamination and exposure of the intrinsic chemical states of BFO. The emergence of Fe$^{2+}$ after etching serves as a clear signature of oxygen vacancy-rich regions within the film. In addition, the appearance of Bi$^{(0)}$ may be attributed to the presence of metallic Bi clusters developed during the ethylene glycol-mediated reduction process.[47] During ambient annealing for crystallization, the BFO surface undergoes oxidation, leading to the characteristic disappearance of Bi$^{(0)}$ and a lower Fe$^{2+}$/Fe$^{3+}$ ratio, which clearly indicates a comparatively reduced oxygen vacancy concentration at the surface. In contrast, depth profiling reveals the appearance of the Bi$^{(0)}$ state along with more than a twofold enhancement in the Fe$^{2+}$/Fe$^{3+}$ ratio, confirming a significantly higher defect density in the subsurface region. These metallic Bi clusters, together with Fe$^{2+}$ centres, promote the formation of localized, percolated conductive pathways throughout the BFO matrix via deep-layer-trapping.[48,49] In its pristine state, BFO behaves as a wide-band-gap insulating oxide due to the low density of electronic states near the Fermi level. However, oxygen vacancy-induced defect states, along with the inherent existence of Bi$^{(0)}$ clusters, introduce energy levels just below the conduction band, effectively reducing the local bandgap and shifting the Fermi level upward.[50] As a result, the electronic energy level of defect-rich BFO becomes closer to that of the top graphene layer, allowing electrons to transfer more easily across the Gr/BFO interface. This favourable alignment enables electron tunnelling through vacancy-assisted channels, producing discrete conductance steps characteristic of quantum transport. Consequently, the Gr/BFO interface acts as a quantum tunnelling junction where defect-mediated conduction, deep layer trapping and interfacial band alignment cumulatively govern the observed quantum conductance behaviour.

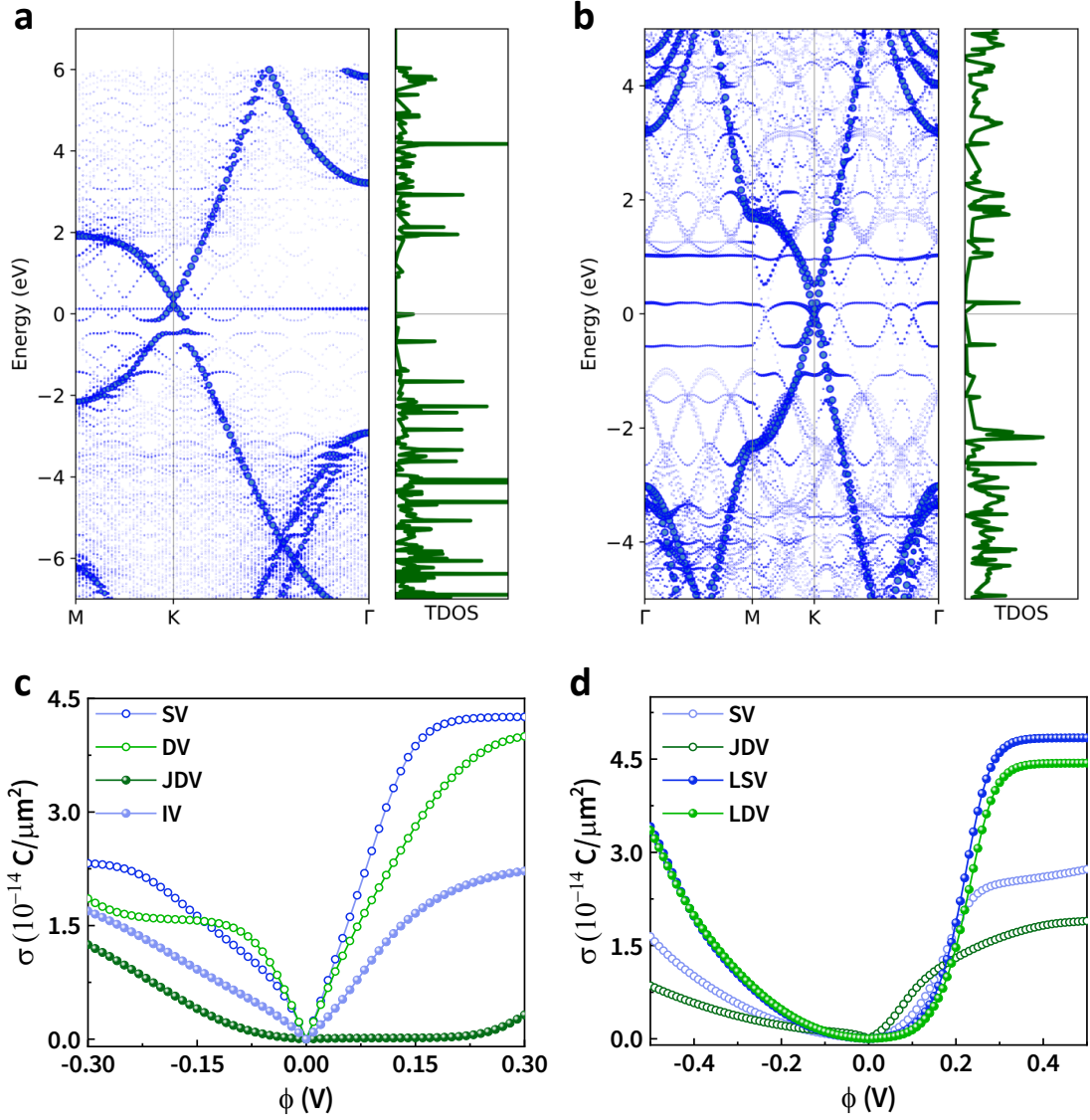

**Figure 4.** *Electronic band dispersion and surface charge density variation in defected graphene. Single point vacancy induced electronic band spectra and total density of states (TDOS) of (a) monolayer (8 × 8 × 1 supercell) and (b) bilayer (5 × 5 × 1 supercell) graphene. The occurrence of defect states at the $E_F$ can be seen, as expected. The response of excess surface charge density with the application of variable external bias in single vacancy (SV), double vacancy (DV), joint double vacancy (JDV), interstitial vacancy (IV) in (c) monolayer graphene and, additional layered single vacancy (LSV), layered double vacancy (LDV) in (d) bilayer graphene.*

**Computational Study of Defect-Induced Charge Localization in Graphene.** To gain quantitative insights into the electronic structure at the graphene (Gr) and validate the role of defect-induced electric field enhancement in driving spatially asymmetric filament formation, we performed comprehensive first-principles density functional theory (DFT) calculations employing a series of point defects in monolayer and bilayer Gr systems, including single vacancy, double vacancy, and interstitial vacancy defects, to model the charge localization behaviour typically observed in Gr. These defects act as trapping centres for excess charge carriers introduced by structural imperfections in the planar carbon network. As a result, a pronounced peak appears in the density of states (DOS) at the Fermi level ($E_F$) (see right panel of Fig. 4a-b), which is otherwise characterized by a dip in pristine Gr.[51–53] Furthermore, as illustrated in the left panels of Fig. 4a–b, the unfolded band dispersions due to single point defects in monolayer and bilayer Gr systems indicate that the defect-induced states reside very close to the Fermi level. This proximity significantly increases the density of localized charges, which in turn generates an induced electric field that regulates the transport behaviour of the underlying BFO layer.[54] The band dispersions due to bivacancy defects in both layered systems are shown in Supplementary Fig. S8. Charge density difference (CDD) analysis further confirms the accumulation of localized charge around the dangling bonds of carbon atoms neighbouring the vacancy sites (Fig. S9 of SI). Notably, the extent of charge localization increases systematically with defect density, particularly in divacancy and interstitial defect configurations. This concentration-dependent charge localization directly explains why multi-layer graphene with inherent defects creates the spatially non-uniform interface conditions necessary for QPC formation, as explained in our proposed mechanism.

Upon applying an external bias voltage ($\varphi$), the Fermi level ($E_F$) of the graphene layer shifts, resulting in the availability of excess charge carriers at the graphene surface. The excess surface charge density ($\sigma$) accumulated in response to applied bias can be quantified via the density of states weighted by the Fermi-Dirac distribution:[55]

$$\sigma = e \int_{-\infty}^{\infty} DOS\ (E)[f_E - f_{E-\varphi}]dE \qquad (1)$$

Where $\varphi$ represents the applied external bias, $f_E$ is the Fermi-Dirac occupation at energy E. At low defect concentrations, the spectral weights of these defect states decrease, while their energetic positions remain nearly unchanged across similar defect types. Figure 4c-d exhibits the variation of $\sigma$ with the application of varied bias voltage at room temperature, 300 K. $\sigma$ responds linearly to modest bias voltages in both defected monolayer and bilayer graphene; as the voltage magnitude increases further, this linear scaling breaks down and $\sigma$ approaches a saturation plateau. In case of single vacancy, the response in $\sigma$ with the external bias is highest, which decreases with the defect density. Since, excessive localization (flat bands) can impede carrier mobility, leading to a degradation of the overall transport properties in the Gr.[56,57] This non-monotonic defect concentration dependence validates that moderate graphene defect densities naturally present in exfoliated ML-Gr provide the optimal balance between electric field localization and carrier availability. This DFT-predicted behaviour aligns well with our mechanistic hypothesis that a strongly localized electric field is generated at graphene interface defect sites with the applied bias.[58] These field-enhanced regions preferentially nucleate oxygen vacancy clustering in the underlying BFO, thereby driving spatially asymmetric filament formation with atomic-scale confinement at the graphene contact.

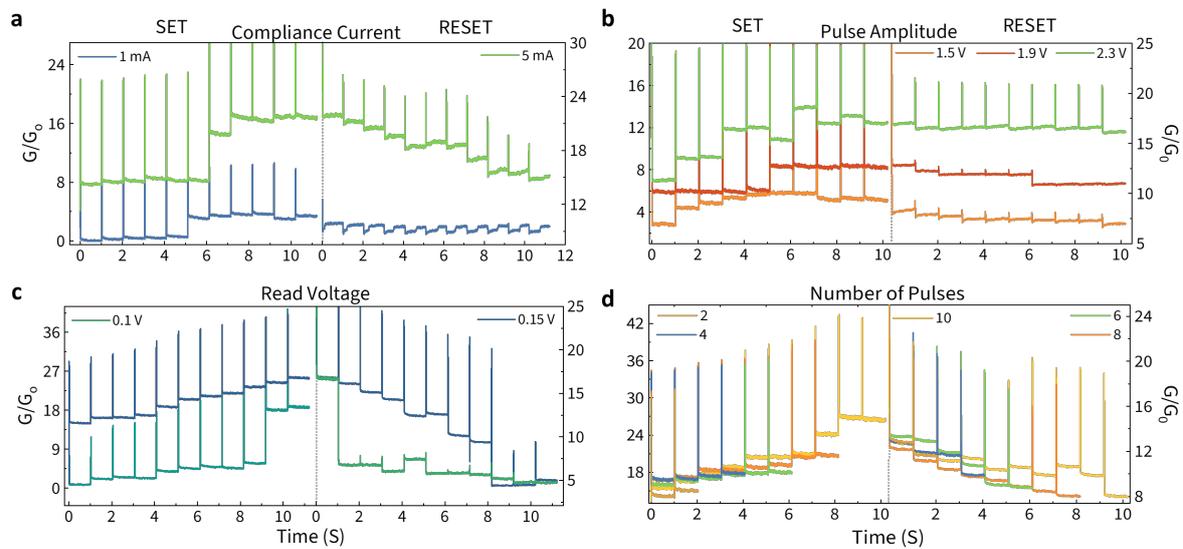

**Figure 5.** *Pulse trains optimization for precise control and stabilization of quantized conductance in Gr/BFO Memristors. Dynamic evolution of quantum conductance states during both SET and RESET process under variations in: (a) compliance current, (b) pulse amplitude, (c) read voltage, (d) number of pulses in the stimulus train.*

**Pulse Driven Quantum Conductance State Stability and Neuromorphic Plasticity.** The ability to precisely control and stabilize discrete conductance levels in ML-Gr/BFO/FTO memristors unveils their promise for neuromorphic applications. By utilizing graphene's exceptional carrier mobility and the tuneable ionic dynamics of the BFO layer on top of the FTO electrode, our devices can carefully translate voltage pulses into well-defined quantum conductance states. In the following pulse-controlled experiments, we demonstrate how variations in compliance current, pulse amplitude, read voltage and pulse count yield stepwise conductance transitions where each plateau represents a potential synaptic weight in an artificial neural network.

Our devices exhibit discrete QC states during both formation (SET) and rupture (RESET) process of CF across multiple pulse configurations, as demonstrated in Fig. 5. The current compliance functions as a protective threshold, preventing permanent dielectric breakdown while governing the peak current magnitude throughout the SET sequences. In Fig. 5a, systematic variation of compliance currents at $1\ and\ 5\ mA$ is performed under increasing pulse amplitude conditions (shown in SI of Fig. S10a) of $pulse\ width\ =\ 20\ ms$, $read\ voltage\ =\ \pm 0.1\ V$ for both SET and RESET operations. It is observed that lower compliance ($1\ mA$) yields only a few narrow conductance steps reaching $\sim 1G_0$ before abrupt RESET, indicating formation of minimal oxygen-vacancy filaments that support lower values quantum conductance channels. Whereas, high compliance ($5\ mA$) shows more successive steps (up to $\sim 20G_0$), reflecting thicker filaments with additional conduction modes and a higher number of quantized plateaus, signifying robust, wide filaments capable of sustaining many parallel quantum conductance channels. This trend confirms that compliance current acts as a "thickness knob" for filament growth, directly setting the maximum number of accessible conductance states by controlling oxygen-vacancy injection and clustering. Figure 5b presents SET/RESET behaviour under three equal pulse amplitudes (shown in SI of Fig. S10b) of $1.5, 1.9\ and\ 2.3\ V$ with constant compliance, $pulse\ width\ =\ 20\ mS, and\ read\ voltage\ =\ \pm 0.1\ V$. These variation in amplitude reveals that low amplitude ($1.5\ V$) yields slow conductance increase and few quantized steps ($\sim 6G_0$), reflecting limited energy to mobilize vacancies. Increasing amplitude ($1.9 - 2.3\ V$) accelerates step formation and elevates final conductance levels ($\sim 12G_0$ at $2.3\ V$), as a higher electric field more efficiently drives filament growth. In the RESET process, same kind of behaviour is observed in the reverse direction. These observations suggest that pulse amplitude directly governs the vacancy drift velocity and filament reinforcement, enabling tuneable state resolution via energy control.[19] Figure 5c shows quantized conductance under read voltages of $0.1\ and\ 0.15\ V$, keeping SET/RESET pulses constant as depicted in Fig. S10c of the SI. Lower read voltages ($0.1\ V$) produce clear, well-defined conductance plateaus with minimal disturbance to the filament. Higher read voltages ($0.15\ V$) lead to elevated baseline conductance and slight filament modification during measurement, resulting in sharp quantized steps in SET and gradual degradation of RESET levels. This indicates that excessive read bias may unintentionally affect filament growth or breakdown, compromising state reliability. Optimal read voltages must therefore balance signal-to-noise ratio with minimal perturbation of quantized states.[59] The effect of conductance state is also observed by varying the number of identical SET/RESET pulses

(2, 4, 6, 8, 10 $pulses$), presented in Fig. 5d. Fewer pulses (2 − 4) yield only initial conductance jumps up to $\sim 18G_0$, as insufficient pulse count limits vacancy accumulation. As we increase the pulse count to 6 − 10, the filament becomes more robust, with conductance plateaus reaching close to $\sim 25G_0$. Similar behaviour appears in the RESET characteristics, which shows that higher SET pulse numbers result in sharp conductance declines during reset, reflecting the presence of larger initial filaments. This indicates that each pulse may slightly modify the filament structure either by injecting more oxygen vacancies or rearranging existing ones. This pulse tunability of QC opens up an extra degree of functionality to mimic synaptic weight update characteristics similar to long-term potentiation (LTP) or long-term depression (LTD) in biological synapses.[38] This pulse-enabled quantization not only reveals the fundamental physics of atomic-scale filament evolution but also establishes a pathway toward hardware that learns and adapts in real time. The presence of controlled bidirectional QC states demonstrated in our ferroelectric oxide system distinguishes this work from the prior reports. Table 1 provides a quantitative comparison between our ML-Gr contacted BFO memristor and established quantum transport devices from the literature.

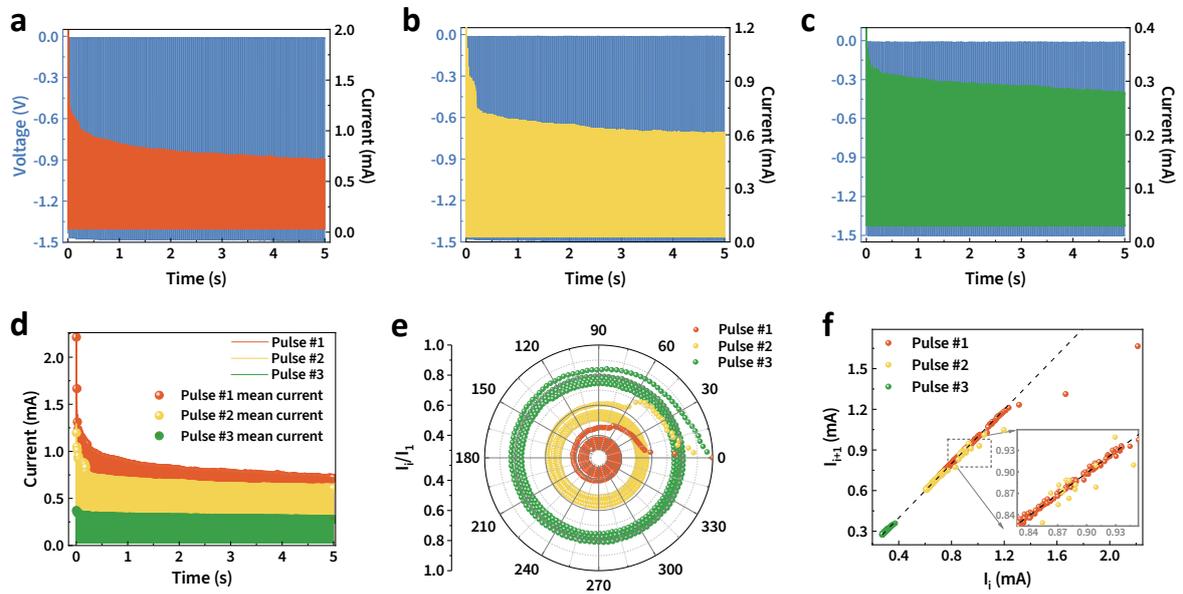

**Figure 6.** *Quantized conductance dynamics during RESET pulse trains in ML-Gr/BFO/FTO memristors. (a–c) Raw Voltage–time traces (blue shading) overlaid with RESET current markers (colour-coded) for successive pulses, illustrating a gradual reduction in peak currents. (d) Recorded current traces for three distinct pulse sequences with corresponding average current values extracted from each pulse. (e) Polar plot of normalized currents ($radius = \frac{I_i}{I_1}$, $angle = (pulse\ number - 1) \times \pi$), with each concentric ring representing a quantized*

*state. (f) Time-lag scatter plot of $I_{i+1}$ versus $I_i$, showing net conductance decrease per pulse (points below dashed unity line) and quantifying cycle-to-cycle variability.*

Quantum conductance levels in memristive filaments should be stable enough at room temperature for their reliable detection.[60] Stability arises from maintaining atomic-scale constrictions against thermal and electrochemical fluctuations, with robust filament compositions such as Magnéli-like nanophases or optimized vacancy distributions.[61] Additionally, random telegraph noise (RTN), caused by stochastic atomic rearrangements near the filament tip, leads to discrete conductance jumps and provides a measure of filament integrity and a source of intrinsic device randomness.[62,63] Understanding these phenomena is key to achieve durable quantized conductance states, essential for multilevel memory and neuromorphic applications.

In the pulse RESET experiment, a train of 500 identical voltage pulses of $P_A = 1.5\,V$ and $P_W = 5\,ms$ is applied to induce the RESET transition as shown in Fig. 6a-c. The first pulse shows the highest current ($I_1$), reflecting the intact conductive channel formed during the SET process. Each subsequent pulses show progressively lower peaks ($I_2, I_3$), as each pulse partially ruptures the filament. The smooth decay in peak values indicates gradual constriction rather than abrupt rupture. To characterize each pulse quantitatively, a representative mean current for each pulse can be obtained by averaging the top portion of each current plateau (shown in SI of Fig. S11); this approach reduces the influence of transient spikes and provides a consistent measure of conductance for quantitative comparison. Figure 6d reveals that overlaying these averaged current points onto the original current-time traces links each quantized level directly to its corresponding segment of the pulse train. This combined view makes it clear which portions of the waveform correspond to stable conductance plateaus. To better visualize these conductance plateaus, the average current of each pulse is normalized to the first-pulse value ($i.e, I_i/I_1$) and plotted over successive pulses as shown in Fig. S12 of SI. These flat regions align with conductance at integer and half-integer multiples of the quantum unit G₀, confirming the presence of atomic-scale constrictions in the filament. A polar plot representation ($radius = i_i/I_1, angle = (pulse\ number - 1) \times \pi$) shown in Fig. 6e further highlights these quantized levels, as each conductance state maps to a concentric ring and each pulse to a radial spoke. This visualization clarifies the stability of each quantum level across multiple pulses. To assess cycle-to-cycle variability, a plot of $I_{i+1}$ versus $I_i$ for consecutive pulses produces a time-lag scatter plot as depicted in Fig. 6f. All points fall below the unity-slope line, confirming net conductance reduction per pulse, while the vertical scatter around each discrete level quantifies stochastic fluctuations in atomic rearrangements. Typical variability ranges of 5–10% reflect the inherent randomness of filament rupture.[64] Together, these analyses demonstrate that ML-Gr/BFO/FTO devices exhibit gradual, controllable RESET dynamics via atomic-scale conductance steps. Such insights are essential for designing reliable multilevel memory and neuromorphic circuits, where precise control over conductance states and an understanding of variability directly impact programming strategies.

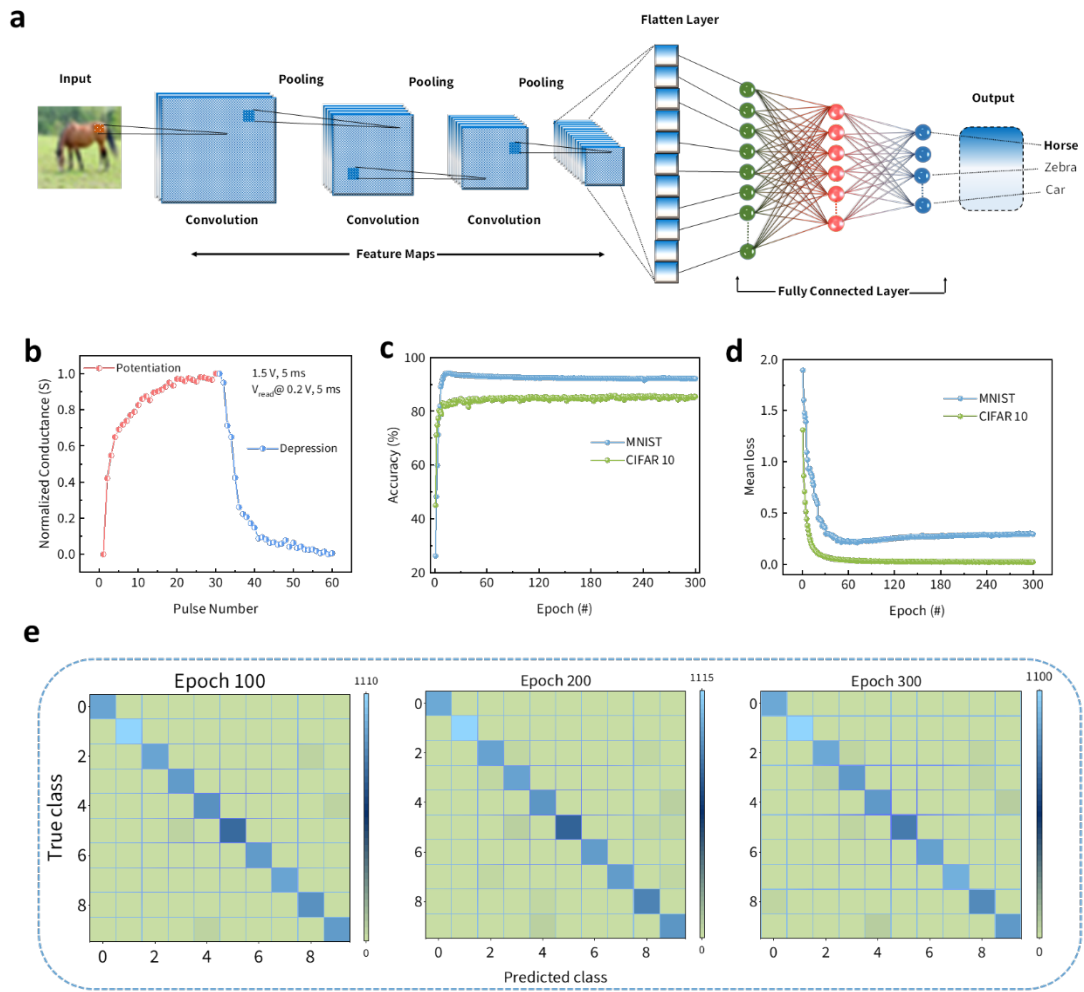

**Figure 7.** *Synaptic plasticity and neuromorphic computing demonstration. (a) Schematic of a Convolutional neural network model deployed to validate the pattern classification capability of the device. (b) Conductance tuning via potentiation and depression under a train of 60 pulses. (c) Progression of pattern classification accuracy, (d) Mean loss progression throughout the 300 training epochs utilizing MNIST and CIFAR-10 datasets. (e)Confusion matrices at training epoch 100. 200 and 300.*

The demonstration of neuromorphic functionality in memristive devices requires comprehensive validation through both device-level characterization and system-level performance evaluation. Figure 7 illustrates the integration of pulse-controlled conductance modulation into a convolutional neural network (CNN) architecture for image classification. The device's pattern recognition capabilities are evaluated using a CNN featuring two convolutional layers followed by two pooling layers, as shown in Figure 7a. Fig. 7b depicts the modulation of the memristor's conductance states in response to a series of voltage pulses. During potentiation, 30 successive positive pulses ($P_A = 1.5\,V, P_w = 5\,ms, V_{read} = 0.2\,V$) gradually increase the device's conductance along a nonlinear path from low to high values. In contrast, during depression, a sequence of the same 30 negative pulses ($P_A = -1.5\,V, P_w = 5\,ms, V_{read} = -0.2\,V$) steadily lowers the conductance back toward its initial value. The observed asymmetry and nonlinearity in these curves arise from the underlying filament formation and rupture processes in the switching layer. Such adjustable conductance levels act as continuous synaptic weights, supporting fine-grained, gradient-based learning rather than binary weight changes. The device's performance for synaptic weight changes is characterized by its potentiation-depression response, with the non-linearity parameters being 5.08 for potentiation and 4.85 for depression, as shown in Fig. S13 in the SI. The non-linearity factor for both potentiation and depression is being calculated by using the following equations:[65]

$$G_{LTP} = B\left(1 - e^{-\frac{P}{A}}\right) + G_{min} \quad (2)$$

$$G_{LTD} = -B\left(1 - e^{\frac{P-P_{max}}{A}}\right) + G_{max} \quad (3)$$

$$B = \frac{G_{max} - G_{min}}{1 - e^{-\frac{P_{max}}{A}}} \quad (4)$$

Where $G_{min}, G_{max}$ and $P_{max}$ are directly obtained from the experimental measurements, representing the lowest conductance, highest conductance, and the maximum number of pulses required to transition the device between these conductance states, respectively. The variable *A* serves as the key factor that controls the degree of nonlinearity in the weight update process, while *B* is derived as a function of *A* designed to ensure that the fitted model accurately aligns within the range of $G_{min}, G_{max}$ and $P_{max}$. This parameterization enables precise modelling of the device's conductance behaviour under varying pulse conditions. Fig. 7c represents how classification accuracy evolves over 300 training epochs for two standard datasets. For MNIST, accuracy increases sharply in the initial epochs and then levels off near its peak (~ 92%), reflecting the straightforward task of digit recognition. Whereas CIFAR-10 shows a slower, more gradual increase in accuracy (~ 85%) and attains a lower plateau, highlighting the greater challenge of distinguishing between varied coloured images. The average loss curve in Fig. 7d represents the reduction in prediction error as the network learns to minimize classification mistakes across training iterations. The average loss in both the MNIST and CIFAR-10 datasets is 0.02 *and* 0.3, respectively. Initially, both datasets display steep drops in loss, reflecting swift improvement as gradients are large and significant weight updates occur in early epochs. With continued training, these curves level off, indicating the network is nearing stable solutions and further reductions in loss are minimal. Notably, the CIFAR-10 dataset maintains higher loss values over the course of training, highlighting the greater challenge posed by its more complex, multi-class image data compared to the simpler MNIST digit dataset, where categories are more easily distinguishable. Finally, Fig. 7e depicts the confusion matrices at training epochs: 100, 200 and 300. These confusion matrices illustrate how the model's classification performance evolves with different training epochs. At 100 epochs, accuracy is moderate with several misclassifications. By epoch 200, correct classifications improve, and errors reduce. At epoch 300, the accuracy is high, showcasing effective learning and strong classification ability of our ML-Gr/BFO/FTO memristor-based neural network.

**Table 1. Comprehensive investigation of the quantum conductance and neuromorphic properties of ML-Gr/BFO/FTO memristor with the previously reported systems.**

| Device Architecture | Max QC ($G_0$) | Bidirectionality | Compliance current | Operating Voltage (V) | Retention (s) | Image Recognition Accuracy |
|---|---|---|---|---|---|---|
| Pt/HfO$_x$/Tin[27] | 10 | No (RESET Only) | 0.1 – 5 mA | ± 2.5 | --- | --- |
| Ag/TiO$_{2-\Delta x}$/Pt[24] | 25 | No (RESET Only) | 1-3 mA | ± 0.7 | $10^4$ | Yes |
| Cu/SiO$_2$/W[15] | 20 | No (RESET Only) | Tens of μA | ± 0.5 | $10^4$ | --- |
| Nb/ZnO/Pt[18] | 16 | Yes (SET & RESET) | 250 µA | ± 1.5 | --- | --- |
| Au/Ti/h-BN/Gr(E)[34] | 1 | Partial | 10 µA | ± 6 | $10^3$ | --- |
| Ta/TaO$_x$/Pt[19] | 15 | No (RESET Only) | --- | ± 1 | $10^3$ | --- |
| W/CeO$_x$/SiO$_2$/NiSi$_2$[17] | 3 | No (RESET Only) | --- | ± 8.5 | --- | --- |
| Ag/SiO$_2$/Pt[66] | 10 | No (RESET Only) | 500 µA | -1 to + 1.5 | $2 \times 10^3$ | --- |
| Au/Ti/hBN/Au/Ti[32] | 10 | No (RESET Only) | 0.1 mA | -3 to +4 | --- | --- |
| ML-Gr/BFO/FTO* *: This Works | 25 | Yes (SET & RESET) | 1 – 10 mA | ± 3 | $6 \times 10^3$ | Yes MNIST (92%) CIFAR-10 (85%) |

**Conclusion:**

In summary, the investigation of multilayer graphene/BFO/FTO memristors has demonstrated precise and bidirectional control of quantum conductance states, achieving both integer and half-integer multiples of $G_0$ up to $25G_0$. Compliance current modulation and tailored pulse schemes enabled dynamic tuning and stabilization of conductance plateaus, with retention over $10^3$ s and cycle-to-cycle reproducibility. Detailed mechanism analysis-combining polar plots, time-lag correlation, and XPS characterization, revealed that oxygen-vacancy dynamics at the graphene-BFO interface unveil the quantized switching. Furthermore, first-principles computational modelling demonstrates that graphene defects generate localized charge accumulation and strong electric field enhancement at the graphene interface, validating our mechanistic hypothesis and establishes the fundamental relationship between graphene engineering and quantum transport phenomena. Integration of these quantized states as synaptic weights in a convolutional neural network yielded image-recognition accuracy on MNIST (~92%) and CIFAR-10 (~85%) datasets, confirming the device's promise for neuromorphic computing applications. This work establishes ferroelectric BiFeO₃ as a versatile platform for quantum memristive technologies. Advanced material engineering and scaling will further improve state stability and integration compatibility of quantum-level memristive devices for next-generation neuromorphic and edge-AI systems.

**Experimental Methods**

**ML-Gr/BFO/FTO memristors fabrication.** The BFO is fabricated using the sol-gel method on a commercially available FTO substrate (Sigma-Aldrich) as discussed in our previously reported work.[35] Few-layer graphene flakes are mechanically exfoliated from a highly oriented pyrolytic graphite (HOPG) substrate using the scotch tape method on a clean $SiO_2$/Si substrate. Then the $SiO_2$/Si substrate is spin-coated with a thin layer of poly(methylmethacrylate) (PMMA). Next, the PMMA/ML-Gr stack is etched out using a dilute NaOH (1:25) solution. After that, the stack is placed into DI water for ~1 hr and transferred onto the as-grown BFO/FTO thin film. Finally, the whole stack is immersed in acetone to remove the PMMA, leaving behind a clean graphene overlayer on BFO/FTO thin film. Some of the optical images of exfoliated ML-Gr on BFO/FTO thin film are shown in fig. S14 of the SI.

**Material Profiling and Device Performance Studies.** Raman spectroscopy is performed with a Horiba T64000 spectrometer equipped with 532 nm excitation at ambient temperature. The parallel-polarized backscattering geometry is selected to optimize signal collection efficiency. High-resolution transmission electron microscopy (HR-TEM) is employed to investigate the fine structure and layered architecture of the BFO thin film. Imaging is performed on a JEOL JEM 2100 instrument operating at an accelerating voltage of 200 kV, which provides sufficient resolution to resolve individual atomic planes and interfacial features. The AFM images are recorded on the Bruker NanoScope Scanning Probe Microscope, and the images are analysed using the Nanoscope analysis 1.5 version software. X-ray photoelectron spectroscopy (XPS) is utilized to determine the elemental composition and bonding environment within the material. A Thermo Fisher Scientific K-Alpha Plus system equipped with monochromatic Al $K_\alpha$ radiation (photon energy 1486.6 eV) is employed for all photoelectron measurements. The experiments are conducted in an ultrahigh vacuum chamber maintained at pressures below $7 \times 10^{-9}$ Torr to minimize sample contamination and ensure reliable quantitative analysis. A Bruker ADVANCE ecoD8 diffractometer equipped with Cu $K_\alpha$ X-ray radiation (wavelength 1.5406 Å) is used as the primary analytical tool for XRD measurement. The instrument is configured at an accelerating voltage of 40 kV, and diffraction data are collected over the angular range from 20° to 80° in 2θ. I-V measurements are performed using a Keithley 4200A-SCS semiconductor parameter analyzer augmented with a 4255-PMU unit designed for high-speed pulse waveform analysis. The integrated system provided comprehensive electrical characterization capabilities spanning DC and dynamic measurement regimes.

**Computational Details.** All first-principles calculations are carried out using density functional theory (DFT) as implemented in the VASP package. The generalized gradient approximation in the Perdew–Burke–Ernzerhof (GGA-PBE) form is employed to treat exchange–correlation effects, while the projector-augmented wave (PAW) method is used to describe electron-ion interactions. Defects are introduced in an $8 \times 8 \times 1$ supercell constructed from monolayer and AB-stacked bilayer graphene. An energy cutoff of 520 eV is chosen for the plane-wave basis set, ensuring well-converged total energies with an electronic self-consistency criterion of $10^{-6}$ eV. Structural optimizations are performed until the residual force on each atom is less than 0.01 eV/Å. Brillouin-zone integrations are carried out using Γ-centred k-point meshes of $19 \times 19 \times 1$ for the primitive cell and $3 \times 3 \times 1$ for the supercell.

**CNN Simulation.** We implemented a convolutional neural network to evaluate pattern recognition capabilities across diverse visual datasets. The model is tested on two complementary datasets: MNIST, a standard collection of normalized $28 \times 28$-pixel grayscale handwritten digits, and CIFAR-10, a more challenging dataset containing $32 \times 32$-pixel colour images representing ten diverse object categories in natural settings. The network architecture proceeded through the following stages: (1) convolutional feature extraction using 16 filters and $3 \times 3$ receptive fields; (2) spatial down sampling through max-pooling with $2 \times 2$ windows, reducing dimensions to $13 \times 13$; (3) additional feature extraction with 32 filters maintaining the same kernel dimensions; (4) further dimensionality reduction through $2 \times 2$ pooling yielding $5 \times 5$ feature maps; (5) flattening of spatial structure; and (6) dense layer classification.

## Supporting Information

Supporting Information is available from the ACS Online Library or from the author.

## Author Contributions

S.R., S.K.M. and S.S. conceived the experiment and designed the study. S.R., S.K.M. and P.S. performed the transport measurement, developed the fabrication method. S.R., H.N.M.,

A.K. and S.S. analyzed the transport measurement and conceived the visuals. S.R. and S.S. modelled the CNN architecture. S.D. and K.G. performed and analyzed the XRD, XPS, AFM, and HRTEM studies, respectively. S.J. and B.R.K.N. performed and analyzed the DFT calculations and wrote the DFT section. S.R., S.K.M. and S.S. wrote and modified the paper, respectively. All authors discussed the results, contributed to the interpretation of data, and contributed to editing the manuscript.

Notes

The authors declare no competing financial interest.

ACKNOWLEDGMENTS

S.S., S.R. and P.S. acknowledge Anusandhan National Research Foundation (ANRF) (CRG/2023/006935) for partial financial support to this work and fellowship, respectively.

**References:**

(1) Kudithipudi, D.; Schuman, C.; Vineyard, C. M.; Pandit, T.; Merkel, C.; Kubendran, R.; Aimone, J. B.; Orchard, G.; Mayr, C.; Benosman, R.; Hays, J.; Young, C.; Bartolozzi, C.; Majumdar, A.; Cardwell, S. G.; Payvand, M.; Buckley, S.; Kulkarni, S.; Gonzalez, H. A.; Cauwenberghs, G.; Thakur, C. S.; Subramoney, A.; Furber, S. Neuromorphic Computing at Scale. *Nature 2025 637:8047* 2025, *637* (8047), 801–812. https://doi.org/10.1038/s41586-024-08253-8.
(2) Upadhyay, N. K.; Jiang, H.; Wang, Z.; Asapu, S.; Xia, Q.; Joshua Yang, J. Emerging Memory Devices for Neuromorphic Computing. *Adv Mater Technol* 2019, *4* (4), 1800589. https://doi.org/10.1002/ADMT.201800589.
(3) Ielmini, D.; Wong, H. S. P. In-Memory Computing with Resistive Switching Devices. *Nature Electronics 2018 1:6* 2018, *1* (6), 333–343. https://doi.org/10.1038/s41928-018-0092-2.
(4) Wang, Z.; Wu, H.; Burr, G. W.; Hwang, C. S.; Wang, K. L.; Xia, Q.; Yang, J. J. Resistive Switching Materials for Information Processing. *Nature Reviews Materials 2020 5:3* 2020, *5* (3), 173–195. https://doi.org/10.1038/s41578-019-0159-3.
(5) Yang, J. J.; Strukov, D. B.; Stewart, D. R. Memristive Devices for Computing. *Nature Nanotechnology 2013 8:1* 2012, *8* (1), 13–24. https://doi.org/10.1038/nnano.2012.240.
(6) Yang, J. J.; Pickett, M. D.; Li, X.; Ohlberg, D. A. A.; Stewart, D. R.; Williams, R. S. Memristive Switching Mechanism for Metal/Oxide/Metal Nanodevices. *Nature Nanotechnology 2008 3:7* 2008, *3* (7), 429–433. https://doi.org/10.1038/nnano.2008.160.

(7) Zhang, Y.; Wang, Z.; Zhu, J.; Yang, Y.; Rao, M.; Song, W.; Zhuo, Y.; Zhang, X.; Cui, M.; Shen, L.; Huang, R.; Joshua Yang, J. Brain-Inspired Computing with Memristors: Challenges in Devices, Circuits, and Systems. *Appl Phys Rev* 2020, *7* (1). https://doi.org/10.1063/1.5124027/997428.

(8) Waser, R.; Aono, M. Nanoionics-Based Resistive Switching Memories. *Nature Materials 2007 6:11* 2007, *6* (11), 833–840. https://doi.org/10.1038/nmat2023.

(9) Waser, R.; Dittmann, R.; Staikov, G.; Szot, K. Redox-Based Resistive Switching Memories – Nanoionic Mechanisms, Prospects, and Challenges. *Advanced Materials* 2009, *21* (25–26), 2632–2663. https://doi.org/10.1002/adma.200900375.

(10) Valov, I.; Lu, W. D. Nanoscale Electrochemistry Using Dielectric Thin Films as Solid Electrolytes. *Nanoscale* 2016, *8* (29), 13828–13837. https://doi.org/10.1039/C6NR01383J.

(11) Milano, G.; Aono, M.; Boarino, L.; Celano, U.; Hasegawa, T.; Kozicki, M.; Majumdar, S.; Menghini, M.; Miranda, E.; Ricciardi, C.; Tappertzhofen, S.; Terabe, K.; Valov, I. Quantum Conductance in Memristive Devices: Fundamentals, Developments, and Applications. *Advanced Materials* 2022, *34* (32), 2201248. https://doi.org/10.1002/ADMA.202201248.

(12) Li, Y.; Long, S.; Liu, Y.; Hu, C.; Teng, J.; Liu, Q.; Lv, H.; Suñé, J.; Liu, M. Conductance Quantization in Resistive Random Access Memory. *Nanoscale Research Letters 2015 10:1* 2015, *10* (1), 1–30. https://doi.org/10.1186/S11671-015-1118-6.

(13) Xue, W.; Gao, S.; Shang, J.; Yi, X.; Liu, G.; Li, R.-W.; Xue, W.; Gao, S.; Shang, J.; Yi, X.; Liu, G.; Li, -W R. Recent Advances of Quantum Conductance in Memristors. *Adv Electron Mater* 2019, *5* (9), 1800854. https://doi.org/10.1002/AELM.201800854.

(14) Terabe, K.; Hasegawa, T.; Nakayama, T.; Aono, M. Quantized Conductance Atomic Switch. *Nature 2004 433:7021* 2005, *433* (7021), 47–50. https://doi.org/10.1038/nature03190.

(15) Nandakumar, S. R.; Minvielle, M.; Nagar, S.; Dubourdieu, C.; Rajendran, B. A 250 MV Cu/SiO2/W Memristor with Half-Integer Quantum Conductance States. *Nano Lett* 2016, *16* (3), 1602–1608. https://doi.org/10.1021/ACS.NANOLETT.5B04296/ASSET/IMAGES/NL-2015-04296V_M003.GIF.

(16) Kharlanov, O. G.; Shvetsov, B. S.; Rylkov, V. V.; Minnekhanov, A. A. Stability of Quantized Conductance Levels in Memristors with Copper Filaments: Toward Understanding the Mechanisms of Resistive Switching. *Phys Rev Appl* 2022, *17* (5), 054035. https://doi.org/10.1103/PHYSREVAPPLIED.17.054035/V1.

(17) Miranda, E.; Kano, S.; Dou, C.; Kakushima, K.; Sué, J.; Iwai, H. Nonlinear Conductance Quantization Effects in CeO x/SiO 2-Based Resistive Switching Devices. *Appl Phys Lett* 2012, *101* (1). https://doi.org/10.1063/1.4733356/126500.

(18) Zhu, X.; Su, W.; Liu, Y.; Hu, B.; Pan, L.; Lu, W.; Zhang, J.; Li, R. W. Observation of Conductance Quantization in Oxide-Based Resistive Switching Memory. *Advanced Materials* 2012, *24* (29), 3941–3946. https://doi.org/10.1002/ADMA.201201506.

(19) Yi, W.; Savel'Ev, S. E.; Medeiros-Ribeiro, G.; Miao, F.; Zhang, M. X.; Yang, J. J.; Bratkovsky, A. M.; Williams, R. S. Quantized Conductance Coincides with State Instability and Excess Noise in Tantalum Oxide Memristors. *Nature Communications 2016 7:1* 2016, *7* (1), 1–6. https://doi.org/10.1038/ncomms11142.

(20) Valov, I.; Linn, E.; Tappertzhofen, S.; Schmelzer, S.; Van Den Hurk, J.; Lentz, F.; Waser, R. Nanobatteries in Redox-Based Resistive Switches Require Extension of Memristor

Theory. *Nature Communications 2013 4:1* 2013, *4* (1), 1–9. https://doi.org/10.1038/ncomms2784.

(21) Yang, Y.; Gao, P.; Li, L.; Pan, X.; Tappertzhofen, S.; Choi, S.; Waser, R.; Valov, I.; Lu, W. D. Electrochemical Dynamics of Nanoscale Metallic Inclusions in Dielectrics. *Nature Communications 2014 5:1* 2014, *5* (1), 1–9. https://doi.org/10.1038/ncomms5232.

(22) Song, M.; Lee, S.; Nibhanupudi, S. S. T.; Singh, J. V.; Disiena, M.; Luth, C. J.; Wu, S.; Coupin, M. J.; Warner, J. H.; Banerjee, S. K. Self-Compliant Threshold Switching Devices with High On/Off Ratio by Control of Quantized Conductance in Ag Filaments. *Nano Lett* 2023, *23* (7), 2952–2957. https://doi.org/10.1021/ACS.NANOLETT.3C00327/ASSET/IMAGES/LARGE/NL3C00327_0005.JPEG.

(23) Banerjee, W.; Hwang, H.; Banerjee, W.; Hwang, H. Quantized Conduction Device with 6-Bit Storage Based on Electrically Controllable Break Junctions. *Adv Electron Mater* 2019, *5* (12), 1900744. https://doi.org/10.1002/AELM.201900744.

(24) Sahu, M. C.; Padhi, P. S.; Mallik, S. K.; Roy, S.; Sahoo, S.; Banik, S.; Sharma, R. K.; Misra, P.; Sahoo, S. Pulse Tunable Quantum Conductance States in Oxygen Vacancy Engineered TiO2−Δx Memristor for Artificial Neural Network Applications. *ACS Appl Electron Mater* 2025, *7* (11), 5271–5281. https://doi.org/10.1021/ACSAELM.5C00694/ASSET/IMAGES/LARGE/EL5C00694_0005.JPEG.

(25) Krishnan, K.; Muruganathan, M.; Tsuruoka, T.; Mizuta, H.; Aono, M. Highly Reproducible and Regulated Conductance Quantization in a Polymer-Based Atomic Switch. *Adv Funct Mater* 2017, *27* (10), 1605104. https://doi.org/10.1002/ADFM.201605104.

(26) Chandrashekar, G.; Madhavanunni Rekha, S.; Bhat, S. V.; Thamankar, R. Quantum Conductance in MoO3/TiO2 Heterojunction Memristors: Crafting Controlled Analog-to-Digital Transition for Multilevel Memory Applications. *ACS Appl Electron Mater* 2025, *7*, 6300. https://doi.org/10.1021/ACSAELM.5C00341/ASSET/IMAGES/LARGE/EL5C00341_0005.JPEG.

(27) Sharath, S. U.; Vogel, S.; Molina-Luna, L.; Hildebrandt, E.; Wenger, C.; Kurian, J.; Duerrschnabel, M.; Niermann, T.; Niu, G.; Calka, P.; Lehmann, M.; Kleebe, H. J.; Schroeder, T.; Alff, L. Control of Switching Modes and Conductance Quantization in Oxygen Engineered HfOx Based Memristive Devices. *Adv Funct Mater* 2017, *27* (32), 1700432. https://doi.org/10.1002/ADFM.201700432.

(28) Jameson, J. R.; Gilbert, N.; Koushan, F.; Saenz, J.; Wang, J.; Hollmer, S.; Kozicki, M.; Derhacobian, N. Quantized Conductance in Ag/GeS 2/W Conductive-Bridge Memory Cells. *IEEE Electron Device Letters* 2012, *33* (2), 257–259. https://doi.org/10.1109/LED.2011.2177803.

(29) Valov, I.; Sapezanskaia, I.; Nayak, A.; Tsuruoka, T.; Bredow, T.; Hasegawa, T.; Staikov, G.; Aono, M.; Waser, R. Atomically Controlled Electrochemical Nucleation at Superionic Solid Electrolyte Surfaces. *Nature Materials 2012 11:6* 2012, *11* (6), 530–535. https://doi.org/10.1038/nmat3307.

(30) Gao, S.; Song, C.; Chen, C.; Zeng, F.; Pan, F. Formation Process of Conducting Filament in Planar Organic Resistive Memory. *Appl Phys Lett* 2013, *102* (14). https://doi.org/10.1063/1.4802092/125089.

(31) Zhang, J.; Li, W.; Mi, J.; -, al; Davies, B.; Szyniszewski, S.; Dias, M. A.; Zhang, Z.; Zhu, X.; Wang, L.; Ye, X.; Gao, R.; Zhang, Y.; Li, R.-W. Quantized Conductance in MoS2 Memristors for High-Accuracy Neuromorphic Computing. *J Phys D Appl Phys* 2025, *58* (21), 215103. https://doi.org/10.1088/1361-6463/ADCFAE.

(32) Roldán, J. B.; Maldonado, D.; Cantudo, A.; Shen, Y.; Zheng, W.; Lanza, M. Conductance Quantization in H-BN Memristors. *Appl Phys Lett* 2023, *122* (20), 203502. https://doi.org/10.1063/5.0147403/2890511.

(33) Liu, D.; Cheng, H.; Zhu, X.; Wang, G.; Wang, N. Analog Memristors Based on Thickening/Thinning of Ag Nanofilaments in Amorphous Manganite Thin Films. *ACS Appl Mater Interfaces* 2013, *5* (21), 11258–11264. https://doi.org/10.1021/AM403497Y/SUPPL_FILE/AM403497Y_SI_001.PDF.

(34) Xie, J.; Patoary, M. N.; Rahman Laskar, M. A.; Ignacio, N. D.; Zhan, X.; Celano, U.; Akinwande, D.; Sanchez Esqueda, I. Quantum Conductance in Vertical Hexagonal Boron Nitride Memristors with Graphene-Edge Contacts. *Nano Lett* 2024, *24* (8), 2473–2480. https://doi.org/10.1021/ACS.NANOLETT.3C04057/ASSET/IMAGES/LARGE/NL3C04057_0004.JPEG.

(35) Roy, S.; Charan Sahu, M.; Kumar Jena, A.; Kumar Mallik, S.; Padhan, R.; Ranjan Mohanty, J.; Sahoo, S.; Roy, S.; Sahu, M. C.; Jena, A. K.; Mallik, S. K.; Padhan, R.; Sahoo, S.; Mohanty, J. R. Analog-Digital Hybridity of Resistive Switching in Ion-Irradiated BiFeO3 Memristor for Synergistic Neuromorphic Functionality and Artificial Learning. *Adv Mater Technol* 2025, *10* (2), 2400557. https://doi.org/10.1002/ADMT.202400557.

(36) Hopkins, P. E.; Adamo, C.; Ye, L.; Huey, B. D.; Lee, S. R.; Schlom, D. G.; Ihlefeld, J. F. Effects of Coherent Ferroelastic Domain Walls on the Thermal Conductivity and Kapitza Conductance in Bismuth Ferrite. *Appl Phys Lett* 2013, *102* (12). https://doi.org/10.1063/1.4798497.

(37) Zhao, J.; Zhou, Z.; Zhang, Y.; Wang, J.; Zhang, L.; Li, X.; Zhao, M.; Wang, H.; Pei, Y.; Zhao, Q.; Xiao, Z.; Wang, K.; Qin, C.; Wang, G.; Li, H.; Ding, B.; Yan, F.; Wang, K.; Ren, D.; Liu, B.; Yan, X. An Electronic Synapse Memristor Device with Conductance Linearity Using Quantized Conduction for Neuroinspired Computing. *J Mater Chem C Mater* 2019, *7* (5), 1298–1306. https://doi.org/10.1039/C8TC04395G.

(38) La Barbera, S.; Vuillaume, D.; Alibart, F. Filamentary Switching: Synaptic Plasticity through Device Volatility. *ACS Nano* 2015, *9* (1), 941–949. https://doi.org/10.1021/NN506735M.

(39) Csonka, S.; Halbritter, A.; Mihály, G.; Jurdik, E.; Shklyarevskii, O. I.; Speller, S.; van Kempen, H. Fractional Conductance in Hydrogen-Embedded Gold Nanowires. *Phys Rev Lett* 2003, *90* (11), 116803. https://doi.org/10.1103/PhysRevLett.90.116803.

(40) Pickett, M. D.; Borghetti, J.; Yang, J. J.; Medeiros-Ribeiro, G.; Williams, R. S.; Pickett, M. D.; Borghetti, J.; Yang, J. J.; Medeiros-Ribeiro, G.; Williams, R. S. Coexistence of Memristance and Negative Differential Resistance in a Nanoscale Metal-Oxide-Metal System. *Advanced Materials* 2011, *23* (15), 1730–1733. https://doi.org/10.1002/ADMA.201004497.

(41) Li, Y.; Long, S.; Lv, H.; Liu, Q.; Wang, W.; Wang, Q.; Huo, Z.; Wang, Y.; Zhang, S.; Liu, S.; Liu, M. Reset Instability in Cu/ZrO2:Cu/Pt RRAM Device. *IEEE Electron Device Letters* 2011, *32* (3), 363–365. https://doi.org/10.1109/LED.2010.2095822.

(42) Kim, D. K.; Suh, D. S.; Park, J. Pulse-Programming Instabilities of Unipolar-Type Niox. *IEEE Electron Device Letters* 2010, *31* (6), 600–602. https://doi.org/10.1109/LED.2010.2045873.

(43) Vicarelli, L.; Heerema, S. J.; Dekker, C.; Zandbergen, H. W. Controlling Defects in Graphene for Optimizing the Electrical Properties of Graphene Nanodevices. *ACS Nano* 2015, *9* (4), 3428–3435. https://doi.org/10.1021/ACSNANO.5B01762.

(44) Wang, X.; Wang, X.; Di, Q.; Zhao, H.; Liang, B.; Yang, J. Mutual Effects of Fluorine Dopant and Oxygen Vacancies on Structural and Luminescence Characteristics of F Doped SnO2 Nanoparticles. *Materials 2017, Vol. 10, Page 1398* 2017, *10* (12), 1398. https://doi.org/10.3390/MA10121398.

(45) Van Ruitenbeek, J.; Masis, M. M.; Miranda, E. Quantum Point Contact Conduction. *Resistive Switching: from Fundamentals of Nanoionic Redox Processes to Memristive Device Applications* 2016, 197–224. https://doi.org/10.1002/9783527680870.CH7.

(46) Xue, W.; Li, Y.; Liu, G.; Wang, Z.; Xiao, W.; Jiang, K.; Zhong, Z.; Gao, S.; Ding, J.; Miao, X.; Xu, X. H.; Li, R. W. Controllable and Stable Quantized Conductance States in a Pt/HfOx/ITO Memristor. *Adv Electron Mater* 2020, *6* (2), 1901055. https://doi.org/10.1002/AELM.201901055.

(47) Zhang, M. L.; Feng, C.; Zhang, W. X.; Luan, X. W.; Jiang, J.; Li, L. F. Synthesis of Bismuth Nanoparticles by a Simple One-Step Solvothermal Reduction Route. *Applied Mechanics and Materials* 2013, *423–426*, 155–158. https://doi.org/10.4028/WWW.SCIENTIFIC.NET/AMM.423-426.155.

(48) Qi, H.; Hu, C.; Wang, Y.; Ali, S.; Hu, J.; Bai, N.; Wang, Q.; Qi, J.; He, D. Percolation Theory Based Model of Conduction Mechanism and Characteristic Contradiction in ZnO RRAM. *Appl Phys Lett* 2021, *119* (21). https://doi.org/10.1063/5.0069763/40721.

(49) Chen, A.; Zhang, W.; Dedon, L. R.; Chen, D.; Khatkhatay, F.; MacManus-Driscoll, J. L.; Wang, H.; Yarotski, D.; Chen, J.; Gao, X.; Martin, L. W.; Roelofs, A.; Jia, Q. Couplings of Polarization with Interfacial Deep Trap and Schottky Interface Controlled Ferroelectric Memristive Switching. *Adv Funct Mater* 2020, *30* (43), 2000664. https://doi.org/10.1002/ADFM.202000664;REQUESTEDJOURNAL:JOURNAL:16163028;WGROUP:STRING:PUBLICATION.

(50) Clark, S. J.; Robertson, J. Energy Levels of Oxygen Vacancies in BiFeO3by Screened Exchange. *Appl Phys Lett* 2009, *94* (2). https://doi.org/10.1063/1.3070532/337154.

(51) Ugeda, M. M.; Fernández-Torre, D.; Brihuega, I.; Pou, P.; Martínez-Galera, A. J.; Pérez, R.; Gómez-Rodríguez, J. M. Point Defects on Graphene on Metals. *Phys Rev Lett* 2011, *107* (11), 116803. https://doi.org/10.1103/PHYSREVLETT.107.116803/SUPP_INFO.PDF.

(52) Nanda, B. R. K.; Sherafati, M.; Popović, Z. S.; Satpathy, S. Electronic Structure of the Substitutional Vacancy in Graphene: Density-Functional and Green's Function Studies. *New J Phys* 2012, *14* (8), 083004. https://doi.org/10.1088/1367-2630/14/8/083004.

(53) Nanda, B. R. K.; Satpathy, S. Strain and Electric Field Modulation of the Electronic Structure of Bilayer Graphene. *Phys Rev B* 2009, *80* (16), 165430. https://doi.org/10.1103/PhysRevB.80.165430.

(54) Bhatt, M. D.; Kim, H.; Kim, G. Various Defects in Graphene: A Review. *RSC Adv* 2022, *12* (33), 21520–21547. https://doi.org/10.1039/D2RA01436J.

(55) John, D. L.; Castro, L. C.; Pulfrey, D. L. Quantum Capacitance in Nanoscale Device Modeling. *J Appl Phys* 2004, *96* (9), 5180–5184. https://doi.org/10.1063/1.1803614.

(56) Lherbier, A.; Dubois, S. M. M.; Declerck, X.; Niquet, Y. M.; Roche, S.; Charlier, J. C. Transport Properties of Graphene Containing Structural Defects. *Phys Rev B Condens Matter Mater Phys* 2012, *86* (7), 075402. https://doi.org/10.1103/PHYSREVB.86.075402/DELIVERABLE/2077AC77-2B0E-4C84-9162-F5510E3BBF0B.

(57) Wang, H.; Wang, Y.; Bai, B.; Guo, X.; Xue, J. Electronic Transport Properties of Graphene with Stone-Wales Defects and Multiple Vacancy Chains: A Theoretical Study. *Appl Surf Sci* 2020, *531*, 147347. https://doi.org/10.1016/J.APSUSC.2020.147347.

(58) Lee, J.; Yang, K.; Kwon, J. Y.; Kim, J. E.; Han, D. I.; Lee, D. H.; Yoon, J. H.; Park, M. H. Role of Oxygen Vacancies in Ferroelectric or Resistive Switching Hafnium Oxide. *Nano Convergence 2023 10:1* 2023, *10* (1), 55-. https://doi.org/10.1186/S40580-023-00403-4.

(59) Singh, C. P.; Singh, V. P.; Ranjan, H.; Pandey, S. K. Performance Analysis and Read Voltage Optimization of E-Beam Evaporated Amorphous SnO2-Based Cross-Cell Resistive Switching Device. *IEEE Trans Electron Devices* 2023, *70* (12), 6637–6643. https://doi.org/10.1109/TED.2023.3326125.

(60) Wagenaar, J. J. T.; Morales-Masis, M.; Van Ruitenbeek, J. M. Observing "Quantized" Conductance Steps in Silver Sulfide: Two Parallel Resistive Switching Mechanisms. *J Appl Phys* 2012, *111* (1). https://doi.org/10.1063/1.3672824.

(61) Hu, C.; McDaniel, M. D.; Ekerdt, J. G.; Yu, E. T. High ON/OFF Ratio and Quantized Conductance in Resistive Switching of TiO2 on Silicon. *IEEE Electron Device Letters* 2013, *34* (11), 1385–1387. https://doi.org/10.1109/LED.2013.2282154.

(62) Li, X.; Zanotti, T.; Wang, T.; Zhu, K.; Puglisi, F. M.; Lanza, M. Random Telegraph Noise in Metal-Oxide Memristors for True Random Number Generators: A Materials Study. *Adv Funct Mater* 2021, *31* (27), 2102172. https://doi.org/10.1002/ADFM.202102172.

(63) Jiang, H.; Belkin, D.; Savel'Ev, S. E.; Lin, S.; Wang, Z.; Li, Y.; Joshi, S.; Midya, R.; Li, C.; Rao, M.; Barnell, M.; Wu, Q.; Yang, J. J.; Xia, Q. A Novel True Random Number Generator Based on a Stochastic Diffusive Memristor. *Nat Commun* 2017, *8* (1), 1–9. https://doi.org/10.1038/S41467-017-00869-X;SUBJMETA.

(64) Roldán, J. B.; Miranda, E.; Maldonado, D.; Mikhaylov, A. N.; Agudov, N. V.; Dubkov, A. A.; Koryazhkina, M. N.; González, M. B.; Villena, M. A.; Poblador, S.; Saludes-Tapia, M.; Picos, R.; Jiménez-Molinos, F.; Stavrinides, S. G.; Salvador, E.; Alonso, F. J.; Campabadal, F.; Spagnolo, B.; Lanza, M.; Chua, L. O. Variability in Resistive Memories. *Advanced Intelligent Systems* 2023, *5* (6), 2200338. https://doi.org/10.1002/AISY.202200338;JOURNAL:JOURNAL:26404567;WGROUP:STRING:PUBLICATION.

(65) Chen, P. Y.; Lin, B.; Wang, I. T.; Hou, T. H.; Ye, J.; Vrudhula, S.; Seo, J. S.; Cao, Y.; Yu, S. Mitigating Effects of Non-Ideal Synaptic Device Characteristics for on-Chip Learning. *2015 IEEE/ACM International Conference on Computer-Aided Design, ICCAD 2015* 2016, 194–199. https://doi.org/10.1109/ICCAD.2015.7372570.

(66) Milano, G.; Zheng, X.; Michieletti, F.; Leonetti, G.; Caballero, G.; Oztoprak, I.; Boarino, L.; Bozat, Ö.; Callegaro, L.; De Leo, N.; Godinho, I.; Granados, D.; Koymen, I.; Menghini, M.; Miranda, E.; Ribeiro, L.; Ricciardi, C.; Suñe, J.; Cabral, V.; Valov, I. A Quantum Resistance Memristor for an Intrinsically Traceable International System of Units Standard. *Nature Nanotechnology 2025* 2025, 1–7. https://doi.org/10.1038/s41565-025-02037-5.